\documentclass[floafix,prd,preprint,showpacs,showkeys,preprintnumbers,nofootinbib,superscriptaddress,11pt]{revtex4-1}
\usepackage[utf8]{inputenc}
\usepackage[sort&compress]{natbib}
\usepackage{ulem}
\usepackage{bm}
\usepackage{times}
\usepackage{amssymb,amsbsy,amsmath,amsfonts}
\usepackage{graphicx}
\usepackage{epstopdf}
\usepackage{float}
\usepackage{color}
\usepackage{morefloats}
\usepackage{rotating}
\usepackage{srcltx}
\usepackage{slashed}
\usepackage{subfigure}
\usepackage{multirow}
\usepackage{verbatim}
\usepackage{hyperref}
\usepackage{overpic}
\usepackage{tabularx}

\begin{document}

\title{Production rates of $D_{s}^{+}D_{s}^{-}$ and $D\bar{D}$ molecules in  $B$ decays}

\author{Jia-Ming Xie}
\affiliation{School of Physics, Beihang University, Beijing 102206, China}

\author{Ming-Zhu Liu}~\email{zhengmz11@buaa.edu.cn}
\affiliation{School of Space and Environment, Beihang University, Beijing 102206, China}
\affiliation{School of Physics, Beihang University, Beijing 102206, China}

\author{Li-Sheng Geng}~\email{lisheng.geng@buaa.edu.cn}
\affiliation{Peng Huanwu Collaborative Center for Research and Education, Beihang University, Beijing 100191, China}
\affiliation{School of Physics, Beihang University, Beijing 102206, China}
\affiliation{Beijing Key Laboratory of Advanced Nuclear Materials and Physics, Beihang University, Beijing, 102206, China}
\affiliation{School of Physics and Microelectronics, Zhengzhou University, Zhengzhou, Henan 450001, China}

\begin{abstract}
 Motivated by the recently discovery of a charmonium  $X(3960)$  in $B$ decays by the LHCb Collaboration, the likely existence of two  bound/virtual states (denoted by $ X_{s\bar{s}}$ and $ X_{q\bar{q}}$) below the $D_{s}^{+}D_{s}^{-}$ and $\bar{D}D$ mass thresholds has been re-examined recently. In this work,  we employ the effective Lagrangian approach to calculate their production rates   in $B$ decays utilizing triangle diagrams.   Our results show that the production yields of  $B^{+}\to X_{s\bar{s}} K^{+}$ and $B^{+}\to X_{q\bar{q}} K^{+}$
are of the order of $10^{-4}$, in agreement with the relevant experimental data, which indicates that,  if  the  $D_{s}^{+}D_{s}^{-}$ and $\bar{D}D$ bound states indeed exist, they can be detected in $B$ decays. Moreover, we calculate the production rate of $B^{+}\to X(3960) K^{+}$ assuming that $X(3960)$ is a  resonant state of  $D_{s}^{+}D_{s}^{-}$ and find that it is also of the order of $10^{-4}$ but a bit smaller than that as a $D_s^+D_s^-$ bound  state.

\end{abstract}


\maketitle

\section{Introduction}

In 2004,  the Belle Collaboration  observed a state around 3940 MeV in the $J/\psi \omega$ invariant mass distribution of the  $B\to J/\psi \omega K$ decay~\cite{Belle:2004lle}, which was later confirmed by the BaBar Collaboration in the same process but  the mass was determined to be  3915 MeV~\cite{BaBar:2007vxr}. In 2009, the Belle Collaboration  observed a state near 3915 MeV  in the  $\gamma\gamma\to J/\psi\omega$ reaction~\cite{Belle:2009and}. Later the BaBar Collaboration determined the quantum number of this state to be  $J^{PC}=0^{++}$~\cite{BaBar:2012nxg}. In 2020, the LHCb Collaboration observed a similar state $\chi_{c0}(3930)$  in the $D^{+}D^{-}$ mass distribution of the $B^{+}\to D^{+}D^{-}K^{+} $ decay~\cite{LHCb:2020pxc,LHCb:2020bls}.    In the Review of Particle Physics (RPG)~\cite{ParticleDataGroup:2020ssz}, all these states are referred to as  $\chi_{c0}(3915)$ and  viewed as a candidate for the $\chi_{c0}(2P)$ charmonium~\cite{Duan:2020tsx,Duan:2021bna}. 
Very recently, the LHCb Collaboration reported a charmonium state named as $X(3960)$ with $J^{PC}=0^{++}$ in the $D_{s}^{+}D_{s}^{-}$ mass distribution of the $B^{+}\to D_{s}^{+}D_{s}^{-}K^{+} $ decay. Its mass and width are determined to be $m=3955\pm6\pm11$ MeV and $\Gamma=48\pm17\pm10$ MeV~\cite{LHCb:2022vsv,LHCb:2022dvn}.

Given that the mass of $X(3960)$ is very different from that of $\chi_{c0}(3915)$, they are not likely to be the same state.  It is also difficult to interpret $X(3960)$ as a conventional  charmonium  because the mass of $\chi_{c0}(3P)$ (in the quark model) is around 4200 MeV  ~\cite{Barnes:2005pb,Li:2009zu}. In Ref.~\cite{Bayar:2022dqa}, M. Bayar et al.  argued that in the chiral unitary approach there exist two states below the  $D\bar{D}$ and  $D_{s}^{+}D_{s}^{-}$ thresholds, respectively, and the  new $X(3960)$ can be understood as an enhancement in the $D_{s}^{+}D_{s}^{-}$ mass distribution.  In Ref.~\cite{Xin:2022bzt}, Xin et al. interpreted  $X(3960)$ as a  $J^{PC}=0^{++}$  $D_{s}^{+}D_{s}^{-}$ molecule in the QCD sum rules approach.  With  a leading order contact range effective field theory Ji et al. showed that either a bound state or a virtual state below the $D_{s}^{+}D_{s}^{-}$ mass threshold is needed to   describe  the   $D_{s}^{+}D_{s}^{-}$ mass distribution of  the  $B^{+}\to D_{s}^{+}D_{s}^{-}K^{+} $ decay~\cite{Ji:2022uie}. 

The likely existence of  molecules near the $D\bar{D}$ and $D_{s}^{+}D_{s}^{-}$ mass thresholds has been studied in several approaches before the experimental discoveries.  In Ref.~\cite{Prelovsek:2020eiw}, Prelovsek et al. performed lattice QCD simulations of  coupled-channel  $D\bar{D}$ and  $D_{s}^{+}D_{s}^{-}$ interactions and found the existence of two bound states near the $D\bar{D}$ and  $D_{s}^{+}D_{s}^{-}$ mass thresholds.   In Ref.~\cite{Gamermann:2006nm}, Gamermann et al.  obtained a $I(J^{PC})=0(0^{++})$ narrow resonance with a mass of 3719 MeV, which mainly couples to  the $D\bar{D}$  and $D_{s}\bar{D}_{s}$ channels.  In our previous works~\cite{Liu:2019stu,Wu:2020job}, we  investigated the $D\bar{D}$ interaction in the one boson exchange~(OBE) model and found that a large cutoff of $\Lambda=1.415$ GeV is needed to generate a $D\bar{D}$ bound state. On the other hand, in order to reproduce   $X(3872)$ as a $D^{\ast}\bar{D}$ bound state in the OBE model, one only needs a cutoff of $\Lambda=1.01$ GeV. From this, one concludes that the $D\bar{D}$  interaction is less attractive than the $D^{\ast}\bar{D}$ interaction, in agreement with several recent studies~\cite{Liu:2020tqy,Dong:2021juy,Peng:2021hkr}.   

In Refs.~\cite{Wang:2020elp,Deineka:2021aeu}, the authors  described the $D\bar{D}$ mass distribution  of $\gamma\gamma\to D\bar{D}$ and demonstrated the existence of a bound state near the $D\bar{D}$ mass threshold.
In Ref.~\cite{Dai:2015bcc}, considering the  state $X(3720)$ mainly coupled to $D\bar{D}$,  Dai et al.  predicted the $D\bar{D}$ mass distribution of   the $B^{-}\to D^{0}\bar{D}^{0}K^{-}$ process. In Ref.~\cite{Li:2015iga}, assuming  $\chi_{c0}(3915)$ as a $D_{s}^{+}D_{s}^{-}$ bound state, Li et al.  estimated the branching ratio of the  $B^{+} \to \chi_{c0}(3915)K^{+}$ decay to be $6\times 10^{-4}$. 
In this work, we assume that there exist two molecules near the $D\bar{D}$ and $D_{s}^{+}D_{s}^{-}$ mass thresholds,  denoted as $X_{q\bar{q}}$  and $X_{s\bar{s}}$, and  employ the effective Lagrangian approach to investigate their production rates in $B$ decays via the triangle mechanism. Such an approach has been applied to study the production of $P_{c}$ and $P_{cs}$  in the  $\Lambda_{b}\to J/\psi p K$~\cite{Wu:2019rog} and $\Xi_{b}\to J/\psi \Lambda K$ decays~\cite{Lu:2021irg}. The production rates of   $D\bar{D}$ and $D_{s}^{+}D_{s}^{-}$ molecules in  $B$  decays  are helpful to probe the nature of   $X(3960)$ as well as to understand the $D\bar{D}$ and $D_{s}^{+}D_{s}^{-}$  interactions.

This work is organized as follows. We introduce the triangle mechanism for the decays of $B^{+}\to X_{s\bar{s}}K^{+}$ and $B^{+}\to X_{q\bar{q}}K^{+}$ and the effective Lagrangian approach in Sec.~II. Results and discussions are given in Sec.~III, followed by a short summary in the last section.

\section{Theoretical formalism}

\begin{figure}[!h]
\begin{center}

\begin{overpic}[scale=.22]{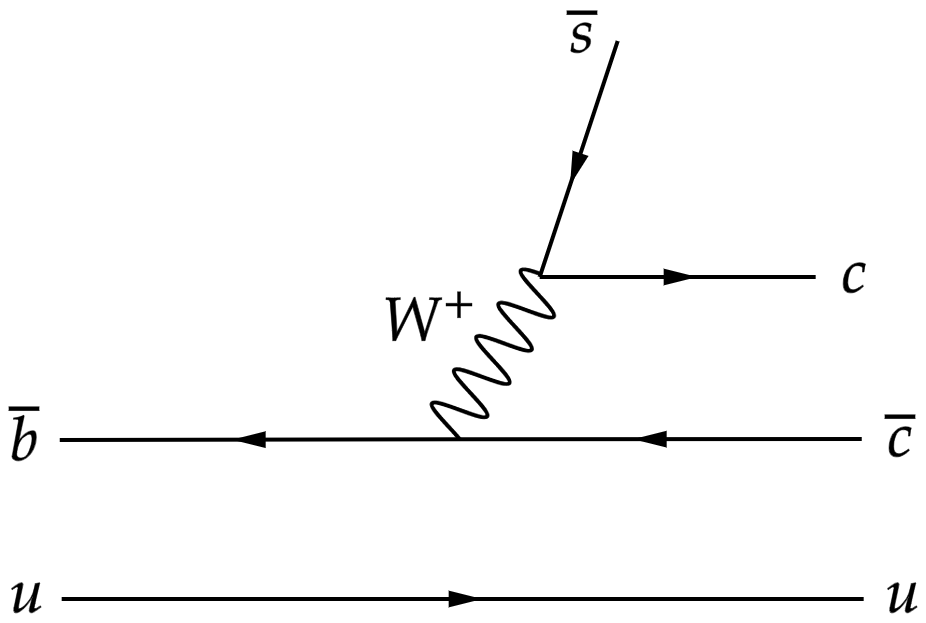}
\end{overpic}
  \caption{ External $W$-emission  for $B^{+}\to D_s^+\bar{D}^{*0}$ and $B^{+}\to D_s^{\ast+}\bar{D}^{0}$.}
  \label{quark}
\end{center}
\end{figure}

\begin{figure}[ttt]
\begin{center}
\begin{overpic}[scale=.24]{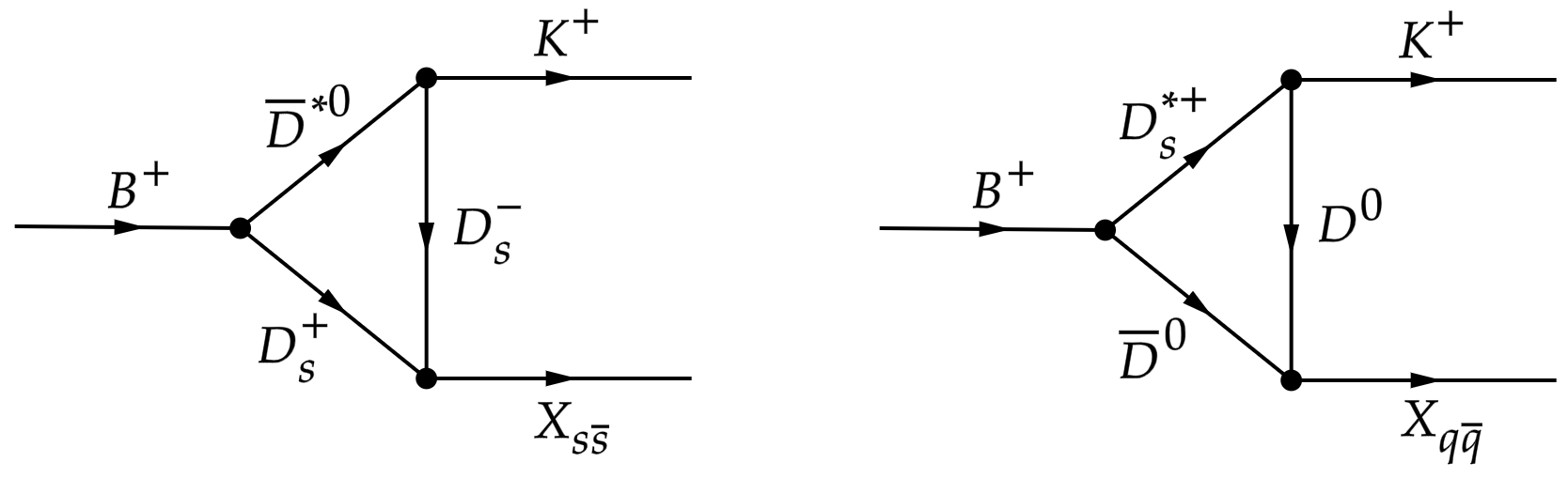}
\end{overpic}
\caption{Triangle diagrams accounting for the two B decays: (a)~$B^{+}\to D_s^+\bar{D}^{*0}\to X_{s\bar{s}}K^{+}$ and  (b)~$B^{+}\to D_s^{\ast+}\bar{D}^{0}\to X_{q\bar{q}}K^{+}$.   }
\label{triangle1}
\end{center}
\end{figure}

The $B$ decays are a good platform to study the weak interaction and  exotic  states~\cite{Cheng:2004ru,Cheng:2008gxa,Chen:2021erj,Wang:2021aql}. In this work, we employ the triangle diagram to investigate the weak decays of $B^{+}\to X_{q\bar{q}} K^{+}$ and  $B^{+}\to X_{s\bar{s}} K^{+}$. 
At the quark level, the decays of $B^{+}\to D_s^+\bar{D}^{*0}$ and $B^{+}\to D_s^{\ast+}\bar{D}^{0}$  can  both proceed through  the external $W$-emission  mechanism shown in Fig.~\ref{quark}.  Referring to the Review of Particle Physics~\cite{ParticleDataGroup:2020ssz}, the absolute branching fractions of the decay modes $B^{+}\to D_{s}^{+} \bar{D}^{\ast0}$ and  $B^{+}\to D_{s}^{\ast+} \bar{D}^{0}$ are $(8.2 \pm 1.7) \times 10^{-3}$  and $(7.6 \pm 1.6) \times 10^{-3}$, respectively, which are larger than those of the $B$ meson  decaying into a charmonium and a $K^{+}$.  
Then, taking into account the interaction vertices of $\bar{D}^{\ast0}\to {D}_{s}^{-}K^{+}$ and $D_{s}^{\ast+}\to D^{0} K^{+}$,   the $D_{s}^{+}D_{s}^{-}$ and $\bar{D}^{0}D^{0}$ molecules can be dynamically generated.  We illustrate the decays of $B^{+}\to X_{s\bar{s}}K^{+}$ and $B^{+}\to X_{q\bar{q}}K^{+}$ at the hadron level via the triangle diagrams  shown in Fig.~\ref{triangle1}. In the following, we introduce the effective Lagrangians relevant to the computation of the Feynman diagrams shown in  Fig.~\ref{triangle1}.   

The  effective Hamiltonian  describing the weak decays of $B^{+}\to D_{s}^{+} \bar{D}^{\ast0}$ and  $B^{+}\to D_{s}^{\ast+} \bar{D}^{0}$ has the following form
\begin{equation}
\mathcal{H}_{eff}=\frac{G_F}{\sqrt{2}}V_{cb}V_{cs}[c_1^{eff}\mathcal{O}_{1}+c_2^{eff}\mathcal{O}_{2}]+h.c.,
\end{equation}
where $G_F $ is the Fermi constant, $V_{bc}$ and $V_{cs}$ are the Cabibbo-Kobayashi-Maskawa~(CKM) matrix elements, $c_{1,2}^{eff}$ are the effective Wilson coefficients, and $\mathcal{O}_{1}$ and $\mathcal{O}_{2}$ are the four-fermion operators of  $(s\bar{c})_{V-A}(c\bar{b})_{V-A}$ and $(\bar{c}c)_{V-A}(s\bar{b})_{V-A}$ with $(\bar{q}q)_{V-A}$ standing for $\bar{q}\gamma_\mu(1-\gamma_5)q$~\cite{https://doi.org/10.48550/arxiv.hep-ph/9806471,Geng:2017esc,Han:2021azw}.

The amplitudes of $B^{+}\to D_{s}^{+} \bar{D}^{\ast0}$ and  $B^{+}\to D_{s}^{\ast+} \bar{D}^{0}$ can be written as the products of two hadronic matrix elements~\cite{Ali:1998eb,Qin:2013tje}
\begin{eqnarray}\label{Ds-KK}
\mathcal{A}\left(B^{+}\to D_{s}^{+} \bar{D}^{\ast0}\right)&=&\frac{G_{F}}{\sqrt{2}} V_{cb}V_{cs} a_{1}\left\langle D_{s}^{+}|(s\bar{c})| 0\right\rangle\left\langle \bar{D}^{\ast 0}|(c \bar{b})| B^{+}\right\rangle\\
\mathcal{A}\left(B^{+}\to D_{s}^{\ast+} \bar{D}^{0}\right)&=&\frac{G_{F}}{\sqrt{2}} V_{cb}V_{cs} a_{1}^{\prime}\left\langle D_{s}^{\ast+}|(s\bar{c})| 0\right\rangle\left\langle \bar{D}^{ 0}|(c \bar{b})| B^{+}\right\rangle
\end{eqnarray}
where $a_{1}^{(\prime)}=c_{1}^{e f f}+c_{2}^{e f f} / N_{c}$  with $N_c=3$ the number of colors. It should be noted that $a_1$ and $a_{1}^{\prime}$ can be  obtained in the factorization approach~\cite{Bauer:1986bm}.

 The current matrix elements between a pseudoscalar meson or vector meson and the vacuum have the following form:
\begin{eqnarray}
\left\langle D_{s}^{+}|(s \bar{c})| 0\right\rangle =g_{\mu \nu} f_{D_{s}^{+}} p^{\nu}_{D_{s}^{+}},   ~~
\left\langle D_{s}^{\ast+} |(s \bar{c} )| 0\right\rangle = m_{ D_{s}^{\ast+}}f_{ D_{s}^{\ast+}}\epsilon_\mu^*,
\end{eqnarray}
where $f_{D_{s}}$  and $f_{ D_{s}^{\ast+}}$ are the decay constants for $D_s$  and $ D_{s}^{\ast+}$, respectively, and $\epsilon_\mu^*$ is the polarization vector of $ D_{s}^{\ast+}$. 
 
The hadronic matrix elements can be parameterised in terms of a few form factors~\cite{Verma:2011yw}
\begin{eqnarray}
&&\left\langle \bar{D}^{\ast0}|(c\bar{b})| B^{+}\right\rangle=\epsilon_{\alpha}^{*}\left\{-g^{\mu \alpha} (m_{\bar{D}^{\ast0}}+m_{B^{+}}) A_{1}\left(q^{2}\right)+P^{\mu} P^{\alpha} \frac{A_{2}\left(q^{2}\right)}{m_{\bar{D}^{\ast0}}+m_{B^{+}}}\right. \\  \nonumber
&&+i \varepsilon^{\mu \alpha \beta \gamma}P_\beta q_\gamma \left.\frac{V\left(q^{2}\right)}{m_{\bar{D}^{\ast0}}+m_{B^{+}}} +q^{\mu} P^{\alpha} \left[\frac{m_{\bar{D}^{\ast0}}+m_{B^{+}}}{q^{2}}A_{1}\left(q^{2}\right)-\frac{m_{B^{+}}-m_{\bar{D}^{\ast0}}}{q^{2}}A_{2}\left(q^{2}\right)-\frac{2m_{\bar{D}^{\ast0}}}{q^{2}}A_{0}\left(q^{2}\right)\right]\right\},
 \\
&&\left\langle \bar{D}^{0}|(c \bar{b} )| B^{+}\right\rangle =\left(p_{B^{+}}+p_{\bar{D}^{0}}\right)^{\mu} F_{1D}(q'^2)+q'^{\mu} F_{2D}(q'^2),
\end{eqnarray}
where $q$ and $q^{\prime}$ represent the momentum of $D_{s}^{+}$ and ${D}_{s}^{\ast+}$, respectively, and $P = (p_{B^{+}} + p_{\bar{D}^{\ast0}})$. The form factors of  $F_{1,2D}(t)$, $A_{0}(t)$, $A_{1}(t)$, $A_{2}(t)$, and $V(t)$ with $t \equiv q^{(\prime)2}$ are parameterized as \footnote{    The electric and magnetic distributions of hadrons in the low energy region, such as those of the nucleons, are often parameterized by dipole form factors of the following form: 
\begin{eqnarray}
G_{E,M}(q^2)=\frac{G_{E,M}(0)}{(1+q^2/m^2)^2},
\end{eqnarray}
which, however, need to be revised in the high energy region~\cite{Arrington:2006zm,Punjabi:2015bba}.  We note that the dipole form factors have also been    adopted to describe the internal structure of baryons in lattice QCD simulations~\cite{Collins:2011mk,Can:2013tna}.  }
\begin{equation}
X(t)=\frac{X(0)}{1-a\left(t / m_{B}^{2}\right)+b\left(t^{2} / m_{ B}^{4}\right)},
\end{equation}
{    which could well fit the transition form factors of $B\to \bar{D}^{(\ast)}$ as shown in Ref.~\cite{Verma:2011yw}.     }

The Lagrangian  describing the interaction between the charmed mesons $D_s$,  $D^*$ and a kaon has the following form  
\begin{eqnarray}
\mathcal{L}_{D_{s} D^{\ast} K}&=& -i g_{D_{s} D^{\ast} K} (D_{s} \partial^{\mu} K D^{\ast\dag}_{\mu}-D_{\mu}^* \partial^{\mu} K  D_{s}^{\dag}), 
\end{eqnarray}
where $g_{D_{s} D^{\ast} K}$ is the coupling constant. 

Assuming that $X_{s\bar{s}}$ and $X_{q\bar{q}}$ are dynamically  generated by the  $D_{s}^{+}D_{s}^{-}$ and $\bar{D}D$  interactions, respectively,   the relevant Lagrangians can be written as 
\begin{eqnarray}
\mathcal{L}_{X_{s\bar{s}} D_{s}^{+}D_{s}^{-}}(x) & = & g_{X_{s\bar{s}} D_{s}^{+}D_{s}^{-}} X_{s\bar{s}}(x) \int dy \Phi(y^2)   D_{s}^{+}(x+\frac{1}{2}y)D_{s}^{-}(x-\frac{1}{2}y), \\
\mathcal{L}_{X_{q\bar{q}}\bar{D}D}(x) & = & g_{X_{q\bar{q}}\bar{D}D} X_{q\bar{q}}(x)\int dy \Phi(y^2) \bar{D}(x+\frac{1}{2}y)D(x-\frac{1}{2}y).
\end{eqnarray}
As  $X_{s\bar{s}}$ and $X_{q\bar{q}}$ are  bound states of $D_{s}^{+}D_{s}^{-}$ and $\bar{D}D$, we adopt the compositeness condition to estimate the couplings of $g_{X_{s\bar{s}} D_{s}^{+}D_{s}^{-}}$ and $g_{X_{q\bar{q}}\bar{D}D}$. 
 The correlation function $\Phi(y^2)$ is introduced to reflect the distribution of the two constituent hadrons in the molecule, which also renders the Feynman diagrams
ultraviolet  finite.  Here we choose the Fourier transformation of the correlation
function in form of a Gaussian function 
\begin{eqnarray}
\Phi(p^2) \doteq \mbox{Exp}(-p_{E}^{2}/\Lambda^2),
\end{eqnarray}
where $\Lambda$ is a size parameter,  which is  expected to be around 1 GeV~\cite{Ling:2021lmq,Ling:2021bir}, and $p_{E}$ is the Euclidean momentum.    The couplings, $g_{X_{s\bar{s}} D_{s}^{+}D_{s}^{-}}$ and $g_{X_{q\bar{q}}\bar{D}D}$, can be estimated by reproducing the binding energies of the $X_{s\bar{s}}$ and $X_{q\bar{q}}$ states via the compositeness condition~\cite{Weinberg:1962hj,Salam:1962ap,Hayashi:1967bjx}.  The condition dictates that  the coupling constant can be determined from the fact that the renormalization constant of the wave function of a composite particle  should be zero. 
The compositeness condition  can be estimated from  the self energy
\begin{equation}
   Z=1-\frac{d \Sigma(k_{0}^{2})}{d{k}_0^{2}}|_{{{k}_0=m_{X}}}=0.
\label{23}
\end{equation}

With the above relevant Lagrangians, one can easily compute  the corresponding amplitudes of Fig.~\ref{triangle1},  
\begin{align}\label{triangleA}
\mathcal{A}_{a} & = g_{X_{s\bar{s}} D_{s}^{+}D_{s}^{-}}\int \frac{d^{4} q_{3}}{(2 \pi)^{4}} \frac{{\rm i}\mathcal{A}(B^{+}\to D_{s}^{+}\bar{D}^{\ast0})\mathcal{A}\left(\bar{D}^{\ast0} \to D_{s}^{-}K^{+}\right)}{\left(q_{1}^{2}-m_{\bar{D}^{\ast}}^{2}+i m_{\bar{D}^{\ast}}\Gamma_{\bar{D}^{\ast}}\right)\left(q_{2}^{2}-m_{D_{s}^{+}}^{2}\right)\left(q_{3}^{2}-m_{D_{s}^{-}}^{2}\right)},  \\
\mathcal{A}_{b} & =g_{X_{q\bar{q}}\bar{D}^{0}D^{0}}\int \frac{d^{4} q_{3}}{(2 \pi)^{4}} \frac{{\rm i}\mathcal{A}(B^{+}\to D_{s}^{\ast+}\bar{D}^{0})\mathcal{A}\left(D_{s}^{\ast+}\to D^{0} K^{+}\right)}{\left(q_{1}^{2}-m_{{D}_{s}^{\ast+}}^{2}+i m_{{D}_{s}^{\ast+}}\Gamma_{{D}_{s}^{\ast+}}\right)\left(q_{2}^{2}-m_{\bar{D}^{0}}^{2}\right)\left(q_{3}^{2}-m_{D^{0}}^{2}\right)}, 
\end{align}
where $q_1$, $q_2$, and $q_3$ denote the momenta of $\bar{D}^{\ast0}$, $D_{s}^{+}$, and $D_{s}^{-}$ for Fig.~\ref{triangle1} (a) and $D_{s}^{\ast+}$, $\bar{D}^{0}$, and $D^{0}$ for Fig.~\ref{triangle1} (b), and $p_{1}$ and $p_{2}$ represent the momenta of $K^{+}$ and $X_{s\bar{s}}(X_{q\bar{q}})$. The vertices of $X_{s\bar{s}} D_{s}^{+}D_{s}^{-}$ and $X_{q\bar{q}}\bar{D}^{0}D^{0}$  are parameterised  as the couplings  
$g_{X_{s\bar{s}} D_{s}^{+}D_{s}^{-}}$ and $g_{X_{q\bar{q}}\bar{D}^{0}D^{0}}$, respectively. 
The  $\bar{D}^{\ast0} \to D_{s}^{-}K^{+}$ transition is expressed as $\mathcal{A}\left(\bar{D}^{\ast0} \to D_{s}^{-}K^{+}\right)=g_{\bar{D}^{\ast0} D_{s}^{-}K^{+}}p_{1}\cdot \varepsilon(q_{1})$, and the  $D_{s}^{\ast+}\to D^{0} K^{+}$ transition is expressed as  $\mathcal{A}(D_{s}^{\ast+}\to D^{0} K^{+})=g_{D_{s}^{\ast+}  D^{0} K^{+}} p_{1}\cdot \varepsilon(q_{1})$. The  weak decay amplitudes of $B^{+}\to D_{s}^{+} \bar{D}^{\ast0}$ and  $B^{+}\to D_{s}^{\ast+} \bar{D}^{0}$ are written as 
\begin{align}\label{am3}
\mathcal{A}(B\to D_{s}\bar{D}^{\ast})&= \frac{G_{F}}{\sqrt{2}}V_{cb}V_{cs} a_{1} f_{D_{s}}\{-q_{2}\cdot \varepsilon(q_{1})(m_{\bar{D}^{\ast0}}+m_{B^{+}}) A_{1}\left(q_{2}^{2}\right)   \\ \nonumber 
&+(k_{0}+q_{1}) \cdot \varepsilon(q_{1}) q_{2}\cdot (k_{0}+q_{1}) \frac{A_{2}\left(q_{2}^{2}\right)}{m_{\bar{D}^{\ast0}}+m_{B^{+}}} +(k_{0}+q_{1}) \cdot \varepsilon(q_{1}) \\ \nonumber       & 
[(m_{\bar{D}^{\ast0}}+m_{B^{+}})A_{1}(q_{2}^2) -(m_{B^{+}}-m_{\bar{D}^{\ast0}})A_{2}(q_2^2) -2m_{\bar{D}^{\ast0}} A_{0}(q_{2}^2)]  \} , \\ \nonumber
\mathcal{A}(B\to D_{s}^{\ast}\bar{D})&=\frac{G_{F}}{\sqrt{2}}V_{cb}V_{cs} a_{1}^{\prime}f_{D_{s}^*}m_{D_{s}^*} \varepsilon(q_{1})\cdot (k_{0}+q_{2}) F_{1D}(q_{1}^2),
\end{align}
 
 With the $B^{+}\to X_{s\bar{s}}(X_{q\bar{q}}) K^{+}$ amplitudes determined  above, 
 the corresponding partial decay widths  can be finally written as
 \begin{eqnarray}
\Gamma=\frac{1}{2J+1}\frac{1}{8\pi}\frac{|\vec{p}|}{m_{B}^2}\bar{|\mathcal{M}|}^{2},
\end{eqnarray}
where $J$ is the total angular momentum of the initial $B$ meson, the overline indicates the sum over the polarization vectors of final states, and $|\vec{p}|$ is the momentum of either final state in the rest frame of  the $B$ meson.

\section{Results and Discussions}\label{sec:Results}

\begin{table}[!h]
\caption{Masses and quantum numbers of  mesons relevant to the present work~\cite{ParticleDataGroup:2020ssz}. \label{mass}}
\begin{tabular}{ccc|ccc}
  \hline\hline
   Meson & $I (J^P)$ & M (MeV) &    Meson & $I (J^P)$ & M (MeV)   \\
  \hline
    $B^{+}$ & $\frac{1}{2}(0^-)$ & $5279.34$ &  $K^{+}$ & $\frac{1}{2}(0^-)$ & $493.677$  \\
   $D^{0}$ & $\frac{1}{2}(0^-)$ & $1864.84$  &    $D^{+}$ & $\frac{1}{2}(0^-)$ & $1869.66$ \\
  $D^{\ast0}$ & $\frac{1}{2}(1^-)$ & $2006.85$ &  $D^{\ast+}$ & $\frac{1}{2}(1^-)$ & $2010.26$   \\
      $D_{s}^{+}$ & $0(0^-)$ & $1968.34$ & 
  $D_{s}^{*+}$ & $0(1^-)$ & $2112.2$ \\
 \hline \hline
\end{tabular}
\label{tab:masses}
\end{table}

The amplitudes of  $B^{+}\to D_{s}^{+} \bar{D}^{0\ast}$ and  $B^{+}\to D_{s}^{\ast+} \bar{D}^{0}$ are obtained by  the naive factorization approach as shown in Eq.~(13). In this work, we take  $G_F = 1.166 \times 10^{-5}~{\rm GeV}^{-2}$, $V_{cb}=0.041$, $V_{cs}=0.987$, $f_{D_{s}} = 250$ MeV, and $f_{D_{s}^*} =272$ MeV  as in Refs.~\cite{ParticleDataGroup:2020ssz,Verma:2011yw,FlavourLatticeAveragingGroup:2019iem,Donald:2012ga,Li:2017mlw}.  
For the form factors, we adopt those of the covariant light-front quark model, e.g.,
  $(F_{1D}(0), a, b)^{B^{+} \to \bar{D}^{0}} = (0.67, 1.22, 0.36)$,  $(A_0(0), a, b)^{B \to \bar{D}^{\ast0}} = (0.68, 1.21, 0.36)$, $(A_1(0), a, b)^{B \to \bar{D}^{\ast0}} = (0.65, 0.60, 0.00)$, and $(A_2(0), a, b)^{B \to \bar{D}^{\ast0}} = ( 0.61, 1.12, 0.31)$~\cite{Verma:2011yw}. Note that the terms containing $V(q^{ 2})$ and $F_{2D}(q^{\prime2})$ do not contribute to the processes we study here. We tabulate the masses and quantum numbers of relevant particles in Table~\ref{tab:masses}. In terms of the branching ratios of $B^{+}\to D_{s}^{+} \bar{D}^{\ast0}$ and $B^{+ }\to D_{s}^{\ast+} \bar{D}^{0} $  we determine  $a_1=0.93$ and $a_{1}^{\prime}=0.81$, consistent with the estimates of Ref.~\cite{Ali:1998eb}. The couplings of $g_{D_{s}^{+}\bar{D}^{\ast0}K^{+}}$ and $g_{D_{s}^{\ast+}\bar{D}^{0}K^{+}}$ are determined by the SU(3)-flavor symmetry, e.g.,  $g_{D_{s}^{+}\bar{D}^{\ast0}K^{+}}=g_{D_{s}^{\ast+}\bar{D}^{0}K^{+}}=\sqrt{2}g_{\bar{D}^{\ast0}\bar{D}^{0}\pi^0}$, where $g_{\bar{D}^{\ast0}\bar{D}^{0}\pi^0}=11.7$ is obtained from the decay width of $D^{*0}\to D^0\pi^0$ \cite{ParticleDataGroup:2020ssz}.
The couplings of $g_{X_{s\bar{s}}D_{s}^{+}D_{s}^{-}}$ and $g_{X_{q\bar{q}}\bar{D}^{0}D^{0}}$ depend on whether the $X_{s\bar{s}}$ and $X_{q\bar{q}}$ states are below or above the mass thresholds of $D_{s}^{+}D_{s}^{-}$ and  $\bar{D}D$, which will be specified below.

 { It is necessary to analyse the uncertainties of our results, which mainly come from the coupling constants  needed to evaluate the triangle diagrams. In the weak interaction vertex,  the uncertainties for the parameters $ X(0) $, $a$ and $b$ of the transition form factors estimated in Ref.~\cite{Verma:2011yw} are quite small and can be neglected. On the other hand, the uncertainties in the experimental  branching ratios of   $B\to \bar{D}^{0}D_{s}^{\ast+}/\bar{D}^{\ast0}D_{s}^{+}$   can propagate to  the parameters $a_1(a_1^{\prime})$ and  lead to about $10\%$  uncertainties for them,   i.e. ,   $a_{1}=0.93^{+0.09}_{-0.10}$ and $a_{1}^{\prime}=0.81^{+0.08}_{-0.09}$.    For the scattering vertices $\bar{D}^{(\ast)}D_{s}^{(\ast)}K$, the couplings  $g_{\bar{D}^{(\ast)}D_{s}^{(\ast)}K}$ are derived via SU(3)-flavor symmetry.   Since  SU(3)-flavor symmetry is broken at the level of  $19\%$~\cite{Durr:2016ulb,Miller:2020xhy},  we attribute an uncertainty of $19\%$  to the $g_{\bar{D}^{(\ast)}D_{s}^{(\ast)}K}$ couplings.  For the couplings $g_{X_{q\bar{q}} D
\bar{D} } $ and $ g_{X_{s\bar{s}} D_s^+ D_s^- } $, we take  a $10\%$ uncertainty. \footnote{ Following Refs.~\cite{Faessler:2007gv,Faessler:2007cu}, we vary the cutoff from 1 GeV to 2 GeV, and the  couplings $g_{X_{q\bar{q}} D
\bar{D} } $ and $ g_{X_{s\bar{s}} D_s^+ D_s^- } $  decrease by approximately $10\%$. Similarly, with the cutoff $\Lambda=2$ GeV and the contact potentials $C_{a}=-5.25$ GeV$^{-2}$ and  $C_{b}=-42.05$ GeV$^{-4}$, we obtain the coupling $ g_{X_{s\bar{s}} D_s^+ D_s^- }=8.69 $~GeV. Compared with the couplings obtained with the cutoff $\Lambda=1$ GeV (as shown later), we find  that the coupling $ g_{X_{s\bar{s}} D_s^+ D_s^- }$ increases by about $10\%$.   }    In terms of the average of the three uncertainties mentioned above,  we arrive at  an uncertainty of $\delta=13\%$ for the couplings characterizing the triangle diagrams. Following Ref.~\cite{Ling:2021asz}, we estimate the uncertainty for the  branching ratios in the following way ${\Gamma}={\Gamma}(1+\delta)^2$. As a result, the calculated branching ratios  have an uncertainty of $28\%$.     }

\begin{figure}[!h]
\centering
\includegraphics[width=10cm]{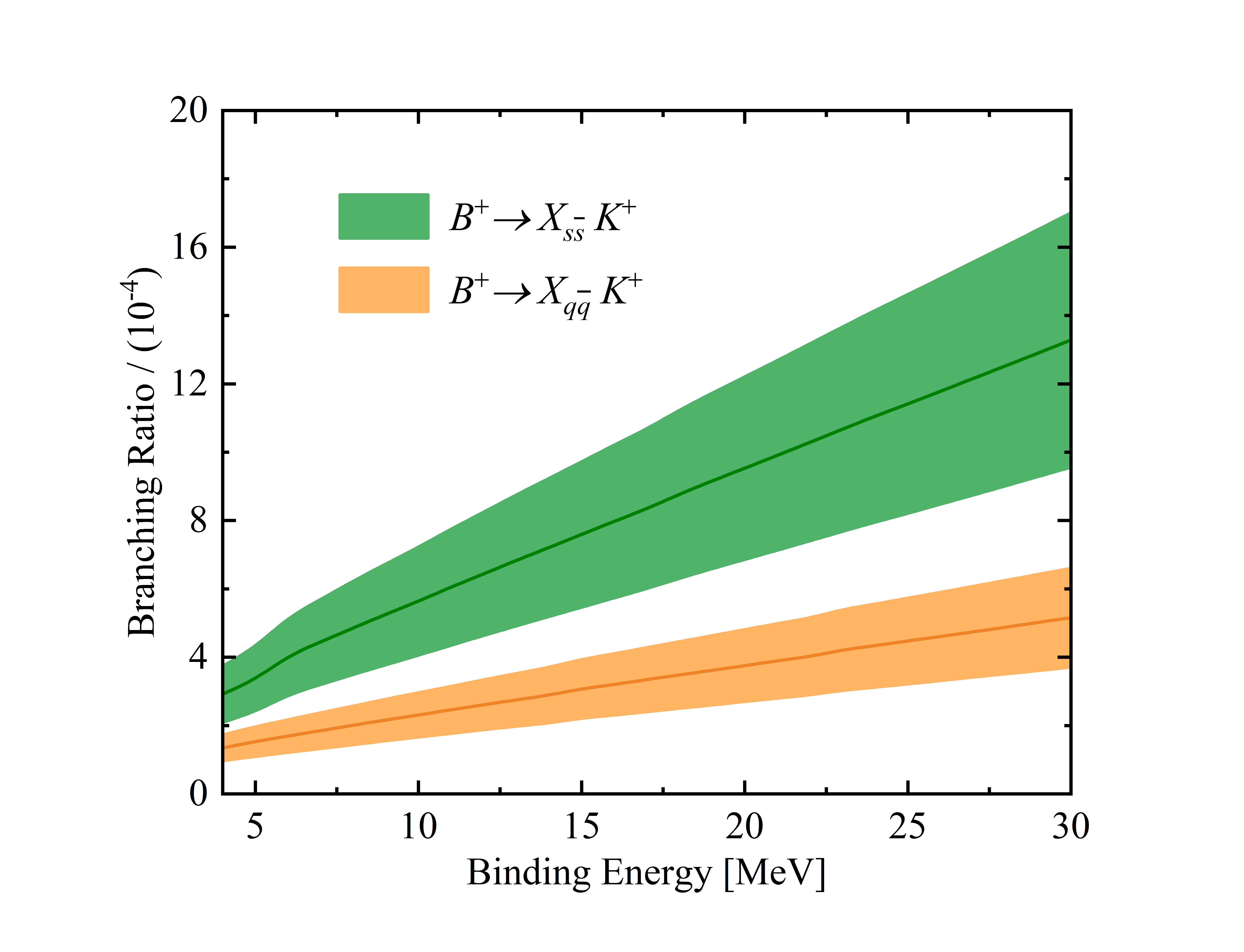}
\caption{  Branching ratios of $B^{+}\to X_{s\bar{s}}K^{+}$ and $B^{+}\to X_{q\bar{q}}K^{+}$ as functions of the binding energies of $D_{s}^{+}D_{s}^{-}$ and $\bar{D}D$ bound states }
\label{Br}
\end{figure}

If $X_{s\bar{s}}$ and  $X_{q\bar{q}}$ are bound states, the coupling of $g_{X_{s\bar{s}}D_{s}^{+}D_{s}^{-}}$ and $g_{X_{q\bar{q}}\bar{D}^{0}D^{0}}$ can be estimated by the compositeness condition~\cite{Ling:2021bir}. According to Refs.~\cite{Prelovsek:2020eiw,Bayar:2022dqa,Ji:2022uie,Dai:2015bcc,Li:2015iga}, we assume that   $X_{s\bar{s}}$ and  $X_{q\bar{q}}$ are located below the mass thresholds of $D_{s}^{+}D_{s}^{-}$ and  $\bar{D}D$ respectively by 4 MeV to 30 MeV. 
  For the $D_{s}^{+}D_{s}^{-}$ state,    the coupling of  $g_{X_{s\bar{s}}D_{s}^{+}D_{s}^{-}}$ is found to range from 9.41 GeV to 20.09 GeV, and the corresponding branching ratios of $B^{+}\to X_{s\bar{s}}K^+$ varies from $(2.9{\pm 0.8})\times 10^{-4}$  to $(13.3{\pm 3.7})\times 10^{-4}$ as shown in Fig.~\ref{Br}. In Ref.~\cite{Li:2015iga}, the authors assumed  $\chi_{c0}(3915)$ as a $D_{s}^{+}D_{s}^{-}$ bound state and estimated the branching ratio of $B^{+}\to \chi_{c0}(3915)K^{+} =6\times 10^{-4}$, which agrees with our result [($9.0{\pm2.5})\times 10^{-4}$]  for a  binding energy of $B=20$ MeV approximately.  
Referring to the Review of Particle Physics~\cite{ParticleDataGroup:2020ssz}, the upper limit of the branching ratio of $B^{+}\to \chi_{c0}(3915)K^{+}$ is $2.8\times 10^{-4}$, which is  smaller than (but consistent with) our result.    One should note that a shallow bound state below the  $D_{s}^{+}D_{s}^{-}$ mass threshold is predicted in two recent works with a binding energy of only several MeV~\cite{Bayar:2022dqa,Ji:2022uie}.  Obviously, the production ratio of such a shallow state in the   $B^{+}\to X_{s\bar{s}}K^{+}$ decay should be smaller than that of a deeply bound state, e.g., $\chi_{c0}(3915)$ treated as a $D_s^+D_s^-$ bound state,  but should be of the same order as shown in Fig.~\ref{Br}. Therefore, our results indicate   that the branching ratio of  $B^{+}\to X_{s\bar{s}}K^{+}$ is of the order of $10^{-4}$ if $X_{s\bar{s}}$ is a bound  state of $D_s^+D_s^-$.

For the $\bar{D}D$ bound state, we determine the coupling of $g_{X_{q\bar{q}}\bar{D}D}$ as $9.62\sim 19.38$ GeV. The coupling of  $g_{X_{q\bar{q}}\bar{D}^{0}D^{0}}$ is further determined by the isospin symmetry, i.e., $g_{X_{q\bar{q}}\bar{D}^{0}D^{0}}=\frac{1}{\sqrt{2}}g_{X_{q\bar{q}}\bar{D}D}$. With the couplings so obtained we calculate the branching ratio of  $B^{+}\to X_{q\bar{q}}K^{+}$, which turns out to be in the range of   $(1.3{ \pm 0.4 }\sim 5.2 { \pm 1.5}) \times 10^{-4}$ as shown in Fig.~\ref{Br}. We note that the only allowed strong decay mode of  a $\bar{D}D$ isoscalar molecule   is  $\eta_{c}\eta$, which implies that the branching ratio of  $B^{+}\to (X_{q\bar{q}}\to \eta_{c}\eta) K^{+}$ is   $10^{-4}$, which agrees  well with the  upper limit of  the branching ratio of   $B^{+}\to  \eta_{c}\eta K^{+}$,  $2.2\times 10^{-4}$~\cite{ParticleDataGroup:2020ssz}. Therefore, our result indicates that  the $\bar{D}D$ bound state can be detected in the $\eta_{c}\eta$ mass distribution of the  $B^{+}\to  \eta_{c}\eta K^{+}$ decay.

At last, we study the scenario where  $X_{s\bar{s}}$ is a resonant state of  $D_s^+D_s^-$, which can be identified as the $X(3960)$ state recently discovered by the LHCb Collaboration. Here we assume that $X(3960)$ is dynamically generated by the $D_s^+D_s^-$ interaction. With a contact potential of the form $C_{a}$ + $C_{b}q^{2}$, one can reproduce the mass and width of $X(3960)$ by solving the Lippmann-Schwinger equation~\cite{Zhai:2022ied}, and then obtain the $X(3960)$ coupling to $D_s^+D_s^-$ from the residues of the pole~\cite{Xie:2022hhv}, where $C_{a}$ and $C_{b}$ are two low energy constants, and $q$ is the momentum of $D_s^{\pm}$ in the  center-of-mass system of the $D_s^+D_s^-$ pair. With the experimental mass and width of  $X(3960)$ as input, we obtain  $C_{a}=-5.15$ GeV$^{-2}$ and $C_{b}=-148.55$ GeV$^{-4}$ for a cutoff of $\Lambda=1$ GeV, and then determine the coupling  $g_{X_{s\bar{s}}D_{s}^{+}D_{s}^{-}}=7.87$ GeV. 
 With the so-determined coupling we calculate  the   branching ratio of  $B^{+}\to X_{s\bar{s}}K^{+}$ and obtain $(1.9{  \pm 0.5} )\times 10^{-4}$.

\section{Summary and Discussion}
\label{sum}

The recently discovered $X(3960)$ by the LHCb Collaboration  motivated us to study the $D_{s}^{+}D_{s}^{-}$ and $\bar{D}D$  molecules predicted in a number of recent works. In this work, we assumed that there exist two bound states below the mass thresholds of $D_{s}^{+}D_{s}^{-}$ and $\bar{D}D$, respectively, and studied their production in the $B$  decays via the triangle mechanism. In such a mechanism,  $B^{+}$ first weakly decays into  $D_{s}^{+}\bar{D}^{\ast0}$ and $D_{s}^{\ast+}\bar{D}^{0}$,   the $\bar{D}^{*0}$/$D_s^{*+}$ decays into $D_s^-$/$D^0$ plus a kaon, and then the final state $D_s^+D_s^-$ and $\bar{D}^0 {D}^0$  interactions dynamically generate  the $X_{s\bar{s}}$ and  $X_{q\bar{q}}$ molecules. As for the bound states, we employed the compositeness condition approach to determine their couplings to their constituents. The resonant state of $D_s^+D_s^-$ is dynamically generated in the single-channel approximation, and  the corresponding coupling is determined from  the residues of the pole.  

We employed the effective Lagrangian approach to calculate the branching ratios of $B^{+} \to X_{s\bar{s}} K^{+} $ and $B^{+} \to X_{q\bar{q}} K^{+} $ assuming that $X_{s\bar{s}}$ and $ X_{q\bar{q}}$ are bound states and found that both of them are of the order of $10^{-4}$. Our results indicate that such bound states of $\bar{D}D$ and $D_{s}^{+}D_{s}^{-}$  (if exist) have large production rates in the $B$ decays since they account for a large portion of the relevant experimental data as shown in Table~\ref{results}.  We note that the $\bar{D}D$ bound state is likely to be detected in the $\eta_{c}\eta$ mass distribution of the $B^{+} \to  \eta_{c}\eta K^{+}$  decay since the $\bar{D}D$ bound state only decays into $\eta_{c}\eta$, while the $D_{s}^{+}D_{s}^{-}$ bound state has more decay modes, e.g., $B^{+} \to  \bar{D}D K^{+}$, $B^{+} \to  \eta_{c}\eta K^{+}$, and $B^{+} \to  J/\psi \omega K^{+}$. 
At last,  assuming that the $X(3960)$ state is dynamically generated by the $D_{s}^{+}D_{s}^{-}$  single-channel interaction, we obtained the  branching ratio of $B^{+} \to X(3960) K^{+} =(1.9{  \pm 0.5} ) \times 10^{-4}$, which can help elucidate the nature of $X(3960)$. 

\begin{table}[!h]
\centering
\caption{Branching ratios of $B\to X_{s\bar{s}/q\bar{q}}/X(3960) K^{+}$ where $X_{s\bar{s}/q\bar{q}}$ is a bound state of  $D_s^+ D_s^-$ or $\bar{D}D$. \label{results}
}
\label{results}
\begin{tabular}{c c c c c c c c}
  \hline \hline
    Decay modes    &~~~~ Our results    &~~~~ Exp~\cite{ParticleDataGroup:2020ssz} 
         \\ \hline 
        $B^{+} \to X_{s\bar{s}} K^{+} $   &~~~~ $({ 2.1\sim 17.0})\times 10^{-4}$   &~~~~ $\mathrm{Br}(B^{+}\to \chi_{c0}(3915)K^{+})<2.8\times 10^{-4}$ 
         \\     $B^{+} \to (X_{q\bar{q}} \to \eta_{c}\eta)  K^{+} $    &~~~~ $({ 0.9\sim6.7})\times 10^{-4}$  &~~~~ $\mathrm{Br}(B^{+} \to \eta_{c}\eta K^{+})<2.2\times 10^{-4}$  
                  \\   \hline 
                  $B^{+} \to X(3960) K^{+} $    &~~~~ $<{ 2.4}\times 10^{-4}$  &~~~~ - 
\\
  \hline \hline
\end{tabular}
\end{table}

\section{Acknowledgments}
  This work is supported in part by the National Natural Science Foundation of China under Grants No.11975041,  No.11735003, and No.11961141004. Ming-Zhu Liu acknowledges support from the National Natural Science Foundation of
China under Grant No.12105007 and  China Postdoctoral
Science Foundation under Grants No. 2022M710317, and No.2022T150036.

\bibliography{reference}
\end{document}